\documentclass[useAMS,usenatbib]{mn2e}

\usepackage{epsfig}
\usepackage{color,soul}

\title{Oscillatory behavior of chromospheric fine structures in a network and a semi-active regions}

\author[Z.F. Bostanc\i, A. G\"ultekin and N. Al]
{Zahide Funda Bostanc\i$^{1,2}$\thanks{Corresponding author, E-mail:
fbostanci@sabanciuniv.edu (ZFB); asumang@istanbul.edu.tr (AG); al@istanbul.edu.tr (AN)}, 
Asuman G\"ultekin$^{2}$, Nurol Al$^{2}$ $$ \\
  $^{1}$Sabanc\i~University, Faculty of Engineering and Natural
  Sciences, Orhanl\i ~Tuzla 34956 Istanbul Turkey\\
  $^{2}$Istanbul University, Faculty of Sciences, Department of
  Astronomy and Space Sciences, University 34119 Istanbul, Turkey}

\begin{document}

\date{}

\pagerange{\pageref{firstpage}--\pageref{lastpage}} \pubyear{2002}

\maketitle

\label{firstpage}

\begin{abstract}
  In the present work, we study the periodicities of oscillations in
  dark fine structures using observations of a network and a
  semi-active region close to the solar disk center. We simultaneously
  obtained spatially high resolution time series of white light images
  and narrow band images in the H$\alpha$ line using the 2D
  G\"ottingen spectrometer, which were based on two Fabry-Perot
  interferometers and mounted in the VTT/Observatorio del
  Teide/Tenerife. During the observations, the H$\alpha$ line was
  scanned at 18 wavelength positions with steps of 125\,m\AA. We
  computed series of Doppler and intensity images by subtraction and
  addition of the H$\alpha$\,$\pm$\,0.3\,\AA\ and~$\pm$\,0.7\,\AA\ pairs,
  sampling the upper chromosphere and the upper photosphere,
    respectively.  Then we obtained power, coherence and phase
  difference spectra by performing a wavelet analysis to the Doppler
  fluctuations. Here, we present comparative results of oscillatory
  properties of dark fine structures seen in a network and a
  semi-active region.

\end{abstract}


\section{Introduction}
When observing the quiet solar chromosphere in the H$\alpha$ line,
dark elongated fine structures can be seen especially in the borders
of the supergranular cells (magnetic network). These structures
  are called mottles on the disk and spicules at the limb and they
  seem to originate from network bright points. Mottles form into two
kind of groups, namely rosettes and chains. They have lengths between
$7''-10''$, widths smaller than $1''$ and lifetimes of the order of 10
minutes \citep{beckers63, beckers68, bray74, suematsu95}. It
  has been considered that mottles are the disk counterparts of
  spicules and that the driving mechanism of these structures has the
  same origin \citep{tsiropoula94, suematsu95, christopoulou01,
    hansteen06, rouppe07}. These jet-like features act as channels
  trough which energy and mass are supplied from the solar photosphere
  into the upper solar atmosphere and the solar wind
  \citep{depontieu04, depontieu06, morton12}.

The active solar chromosphere in the H$\alpha$ line core shows another
sort of dark fine features named fibrils. They are mostly classified
into two types, traditional fibrils and dynamic fibrils. Dynamic
fibrils are jet-like features with higher decelerations, slightly
lower velocities, shorter lifetimes and shorter lengths and located
mostly in a dense active region plage, while traditional fibrils are
low-lying features with lower decelerations, slightly higher
velocities, longer lifetimes and longer lengths \citep[for a more
detailed description and their properties of these structures
see][]{hansteen06, depontieu06, tsiropoula12}.

The study of oscillations and periodicities of these structures has
become more important in recent years as they are believed to be an
important source for the heating of the solar chromosphere.  
  Several authors have investigated periodicities and oscillations in
  these structures. Dominant periods of order 5 minutes are reported
in chromospheric mottles \citep{tziotziou04, tsiropoula09,
  bostanci11}. Similarly, the periods for the oscillations in more
inclined dynamic fibrils are found to be in the 4 -- 6 minute range
\citep{depontieu04, hansteen06, depontieu06}. These findings
  support the idea that normally evanescent photospheric 5 minute
  oscillations leak into the higher atmospheric layers along inclined
  flux tubes \citep{depontieu04}.

Observations performed at photospheric levels show power
  enhancements at higher frequencies ($v$\,$>$\,5\,mHz) in the
  surrounding active regions \citep{brown92, braun92, hindman98,
    thomas00, muglach03} and in quiet regions around the network
  magnetic elements \citep{krijger01, muglach05}. Such power
  enhancements are called power halos. On the other hand,
  chromospheric observations show power suppression called magnetic
  shadows at higher frequencies especially around magnetic network
  elements as well \citep{judge01, krijger01, finsterle04,
    reardon09}.  \cite{kontogiannis10} observed in a rosette region
at photospheric and chromospheric heights power enhancement and
suppression for both the 5 and 3 minutes period bands.

Moreover, the transverse oscillations have also been investigated
  in spicules \citep{zaqarashvili09, okamoto11}, in mottles
  \citep{rouppe07, kuridze12, morton12, kuridze13} and in fibrils
  \citep{pietarila11, morton14}. There are several possible
  interpretations of these oscillatory phenomena such as kink waves
  propagating along magnetic flux tubes, volume-filling Alfv{\rm
    $\acute{e}$}n waves propagating in surrounding spicular
  structures, and transverse pulses excited in the photospheric
  magnetic flux tube by means of buffeting of granules \citep[see][and
  references therein]{zaqarashvili09}.

In this paper, taking advantage of 2D spectroscopic H$\alpha$
observations, we investigate the oscillations in chromospheric dark 
fine structures at different levels of height in the solar atmosphere.

\section[]{Observations \& Data reduction}

Observations of a network and a semi-active region were carried out in
H$\alpha$ near the solar disk center in May 2002. The data were
obtained under good seeing conditions with the `G\"ottingen'
Fabry-Perot spectrometer, which was mounted in the Vacuum Tower
Telescope at the Observatorio del Teide/Tenerife \citep{koschinsky01,
  bostanci11}. During the observations, 8 narrow-band images were
recorded at 18 wavelength positions by scanning through the
H$\alpha$ line. A spacing of 125\,m\AA\ between adjoining wavelength
positions was selected for the spectral scans. The exposure time was
30~ms and the time interval between the start of two consecutive
spectral scans was 49~s. The observed field of view of the raw data
was 38\farcs4$\times$28\farcs6 with an image scale of 0\farcs1 per
pixel. Broad-band images were simultaneously taken with narrow-band
images. Totally 60 spectral scans per region were acquired with the
spectrometer. After the correction for dark current, flat field and
image motion, the broad-band images per each scan were restored by
using the spectral ratio \citep{luhe84} and the speckle masking
technique \citep{weigelt77}. The narrow-band images at each
wavelength position sampling H$\alpha$ line were reconstructed by
using \cite{keller92}'s method.

Since our analysis requires the determination of Doppler and intensity
fluctuations at individual height levels of the solar chromosphere,
the H$\alpha$ line profile for each pixel in the field of view (FOV)
was reconstructed from intensity values of narrow band images
belonging to 18 wavelength positions using spline interpolation.

We computed image time series at $\pm$\,0.3\,\AA\ and $\pm$\,0.7\,\AA\
from the H$\alpha$ center so as to sample different layers of the
solar atmosphere. We then formed intensity and Doppler time series by
averaging and subtracting image pairs for each atmospheric
layer. After the alignment of data cubes using Fourier
cross-correlation techniques, the final FOV sizes of the network and
semi-active regions reduced to 28\farcs8$\times$16\farcs2 and
30\farcs6$\times$16\farcs7, respectively.

In order to investigate the temporal properties of both observed
regions, we performed the wavelet analysis to the Doppler fluctuations
using the wavelet package \citep{torrence98} with a Morlet wavelet
function as the analysing function. Furthermore, we calculated the
global wavelet spectrum for each pixel in the FOV of each observed
region to obtain the spatial distribution of power.  We then averaged
the power over the following frequency bands; 0.73\,--\,1.75\,mHz,
2.46\,--\,4.16\,mHz and 4.93\,--\,8.31\,mHz.

We employed a cross wavelet transform \citep{torrence98} to derive
coherence and phase difference spectra between Doppler fluctuations at
H$\alpha$\,$\pm$\,0.3\,\AA\ and H$\alpha$\,$\pm$\,0.7\,\AA\, sampling
the upper chromospheric and the upper photospheric heights,
respectively.  Just as in the same way as the power maps, coherence
and phase differences for each pixel in the FOV were calculated and
then averaged over the selected frequency bands. Since we took
  the difference between chromospheric and photospheric signals,
  positive and negative values in the phase difference maps indicate
  upward and downward propagating waves, respectively.

%--------------------------------------------

\begin{figure}
\centering
\includegraphics[trim=-0.2cm -0.2cm 8cm 0cm, clip=true, totalheight=0.215\textheight, angle=0]{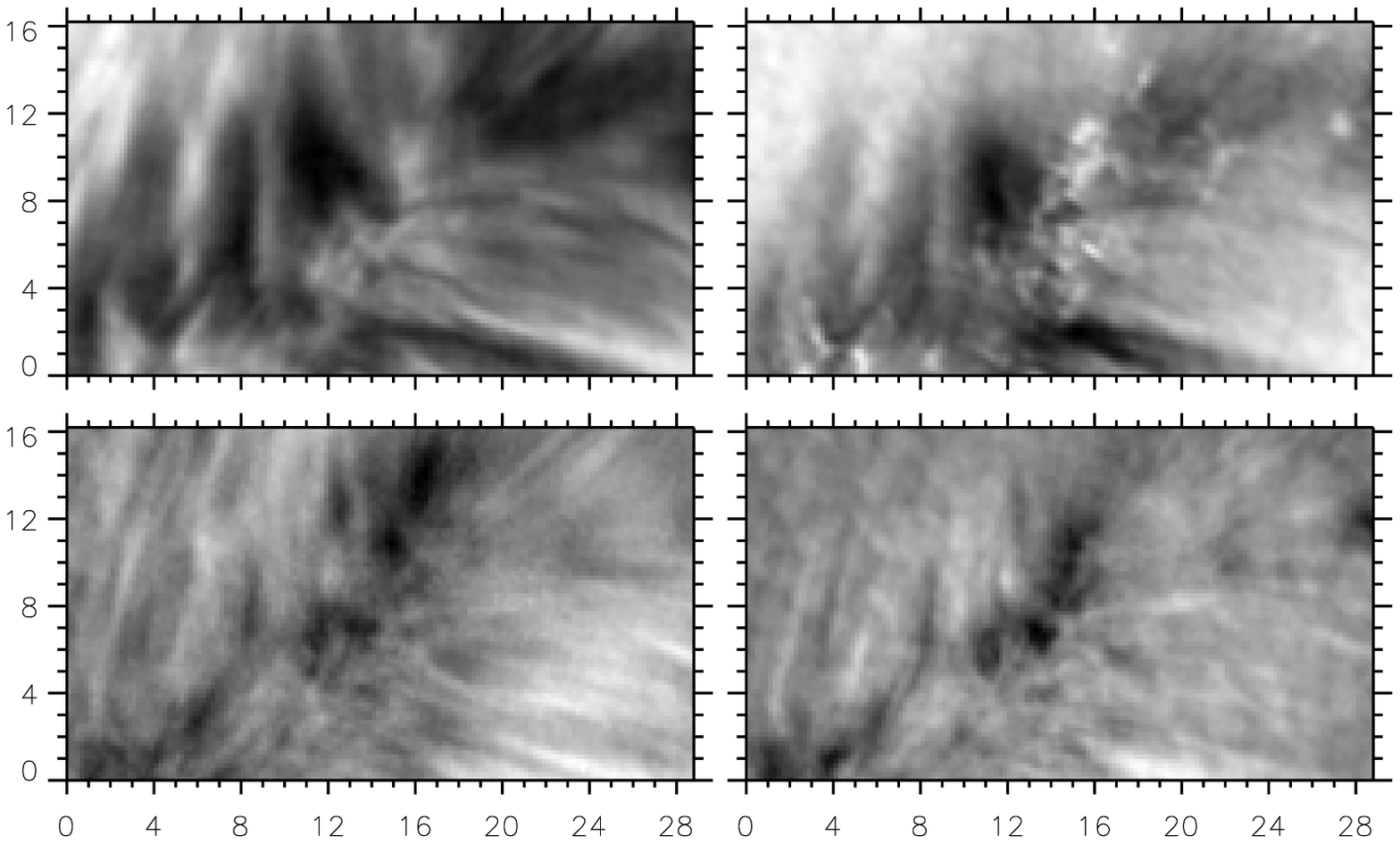}\\
{\large The Network Region}\\
\includegraphics[trim=-0.2cm -0.2cm 8cm 0cm, clip=true, totalheight=0.21\textheight, angle=0]{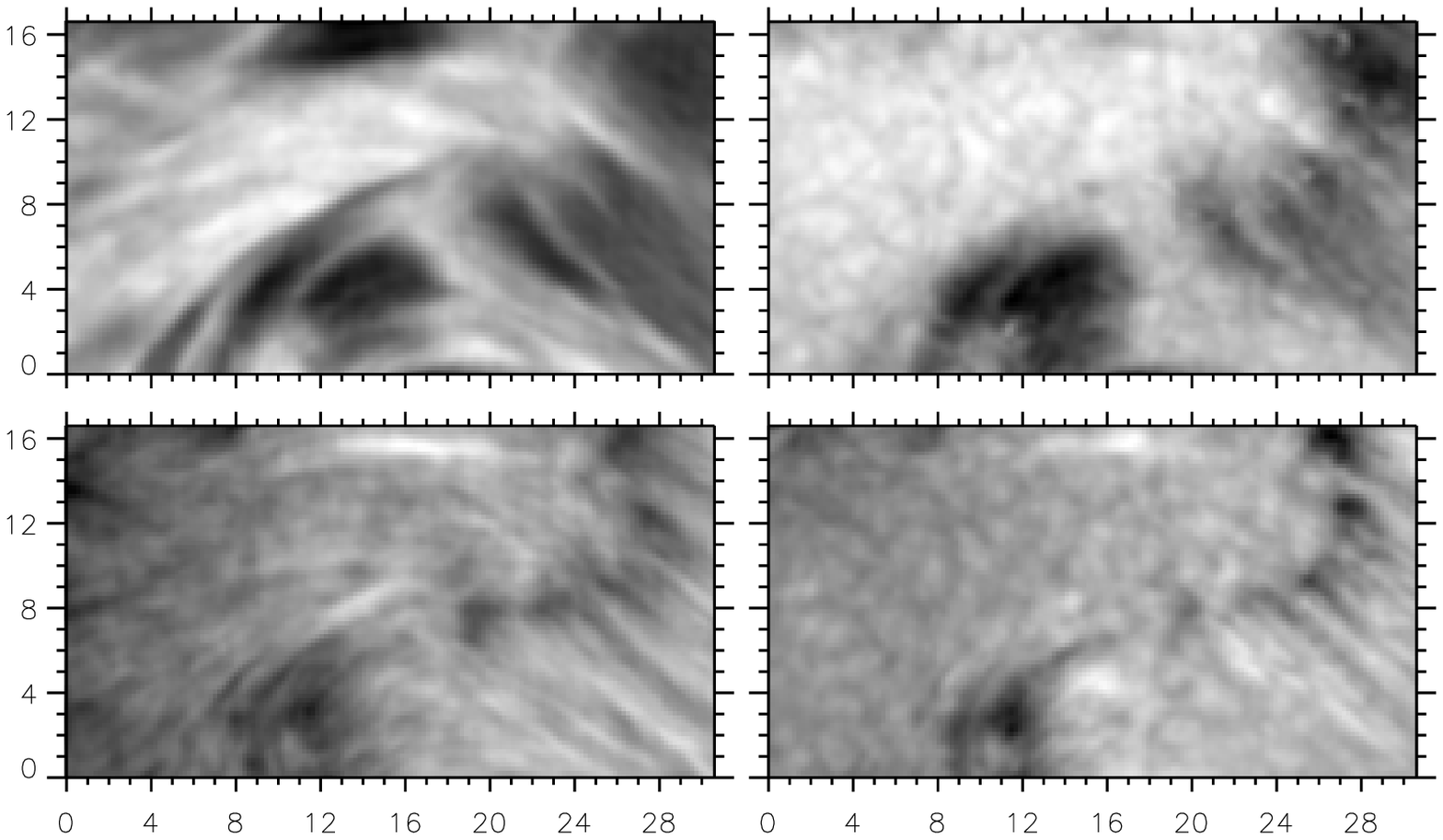}\\
{\large  The Semi-Active Region}
\caption{Temporal averaged intensity ({\it first row of each panel})
  and Doppler images ({\it the second row of each panel}) for two
  different atmospheric depths from the network ({\it upper panel}) and the
  semi-active region ({\it lower panel}). Darker parts in the Doppler
  images indicate downflows while brighter ones upflows. The
  atmospheric depths change from left to right, corresponding to the
  wavelengths at 0.3 and 0.7\,\AA\ from H$\alpha$ center,
  respectively. The sizes of the FOV for the network and semi-active
  regions are 28\farcs8$\times$16\farcs2 and
  30\farcs6$\times$16\farcs7, respectively.}
\label{fig:meanimage}
\end{figure}
%-------------------------------------------- 

\section[]{Results \& Discussions}
We show in Figure~\ref{fig:meanimage} the temporal averaged intensity
and Doppler images corresponding to different depths in the atmosphere
for the network (upper two rows) and the semi-active (lower two rows)
regions. In the network region, mottles spread out from bright points
occupying the network boundary, while in the semi-active region
fibrils extend from two groups of bright points located around the
lower center and the upper right of the FOV. Morphologies of mottles
and fibrils vary in relevant intensity images belonging to different
atmospheric layers. They are highly structured and more elongated in
different inclinations and directions at 0.3\,\AA\ from H$\alpha$
center, corresponding to the chromosphere, but they are less
outstanding at 0.7\,\AA\ from the line center, corresponding to the
upper photosphere \citep[for details on H$\alpha$ line formation see,
e.g.,][]{leenaarts12}. Moreover, the pattern of granules and bright
points become visible at the lowest layer. In the temporal averaged
Doppler images, darker and lighter shades of grey indicate downflows
and upflows, respectively. Brighter and darker elongated structures
seen in areas where mottles and fibrils are found in the intensity
images, point to the presence of upward and downward motion along
these structures. Darker areas close to the bright points indicate
that the downflows are significant at the foot point of mottles and
fibrils.

\begin{figure*}
\begin{center}
\includegraphics[trim=-0.2cm -0.2cm -0.2cm 4.5cm, clip=true, totalheight=0.31\textheight, angle=0]{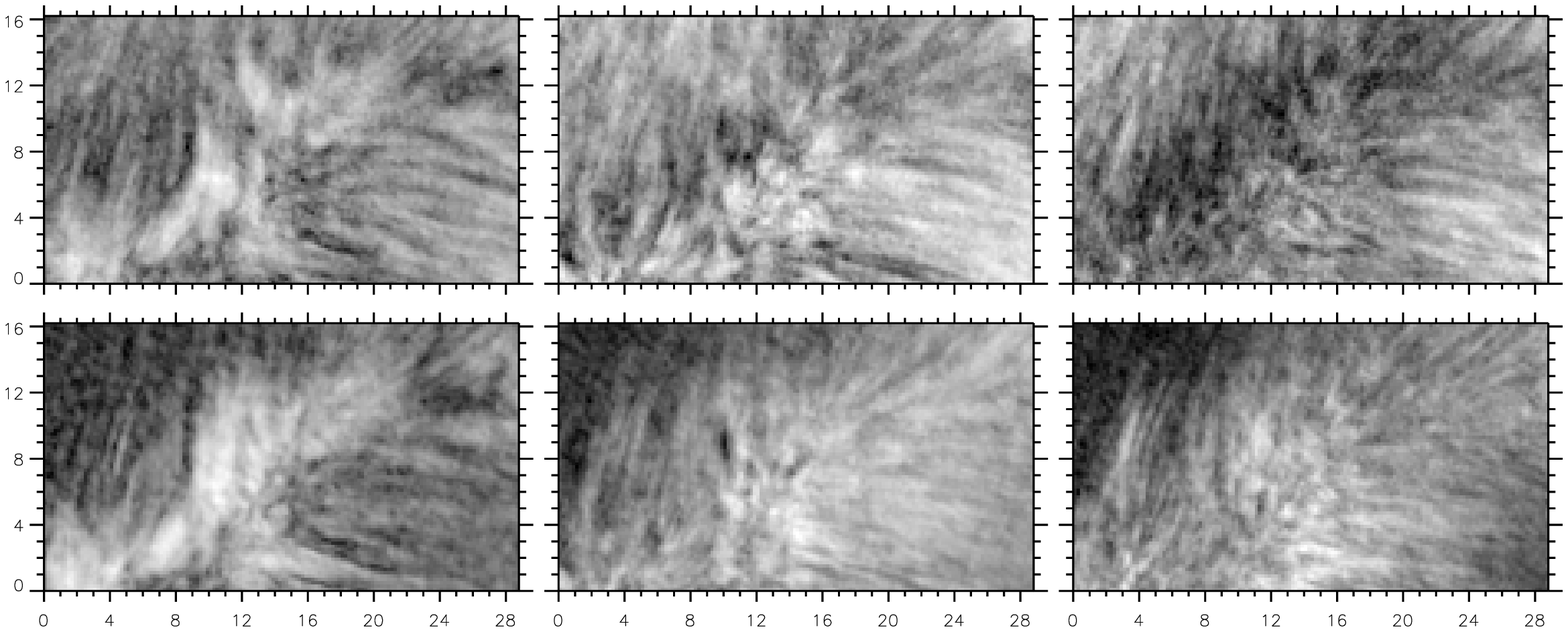}\\
{\large The Network Region}\\
\includegraphics[trim=-0.2cm -0.2cm -0.2cm 4.5cm, clip=true, totalheight=0.3012\textheight, angle=0]{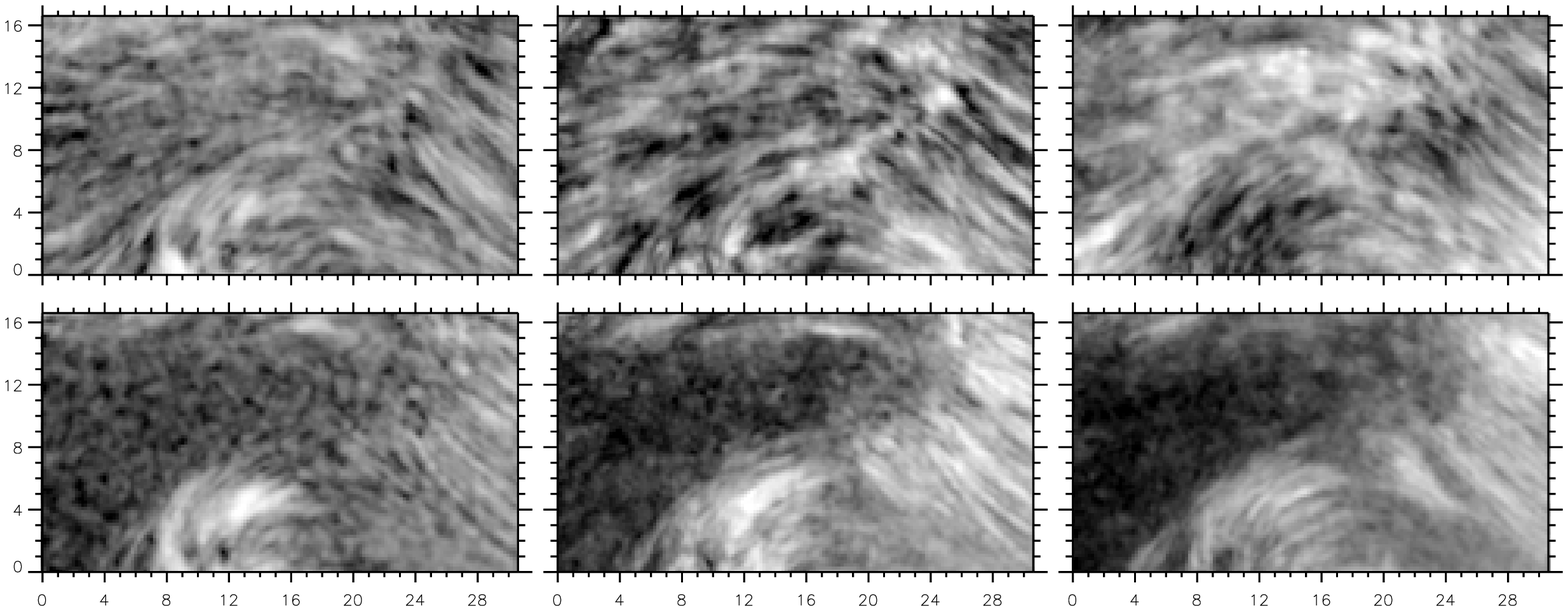}\\
{\large The Semi-Active Region}\\
\caption{Power maps obtained from Doppler image sequences at two
  different atmospheric layers for the network region ({\it upper six
    panels}) and the semi-active region ({\it lower six
    panels}). Columns from left to right correspond to frequency
  ranges of 0.73\,--\,1.75, 2.46\,--\,4.16 and 4.93\,--\,8.31\,mHz,
  respectively. Rows from up to down for each observed region
  correspond to the wavelengths at 0.3 and 0.7\,\AA\ from the line
  center, respectively. The intensity scale is logarithmic and
  brighter patches show strong power areas.}
\label{fig:2D_fourier}
\end{center}
\end{figure*}

\begin{figure*}
\begin{center}
\includegraphics[scale=0.71]{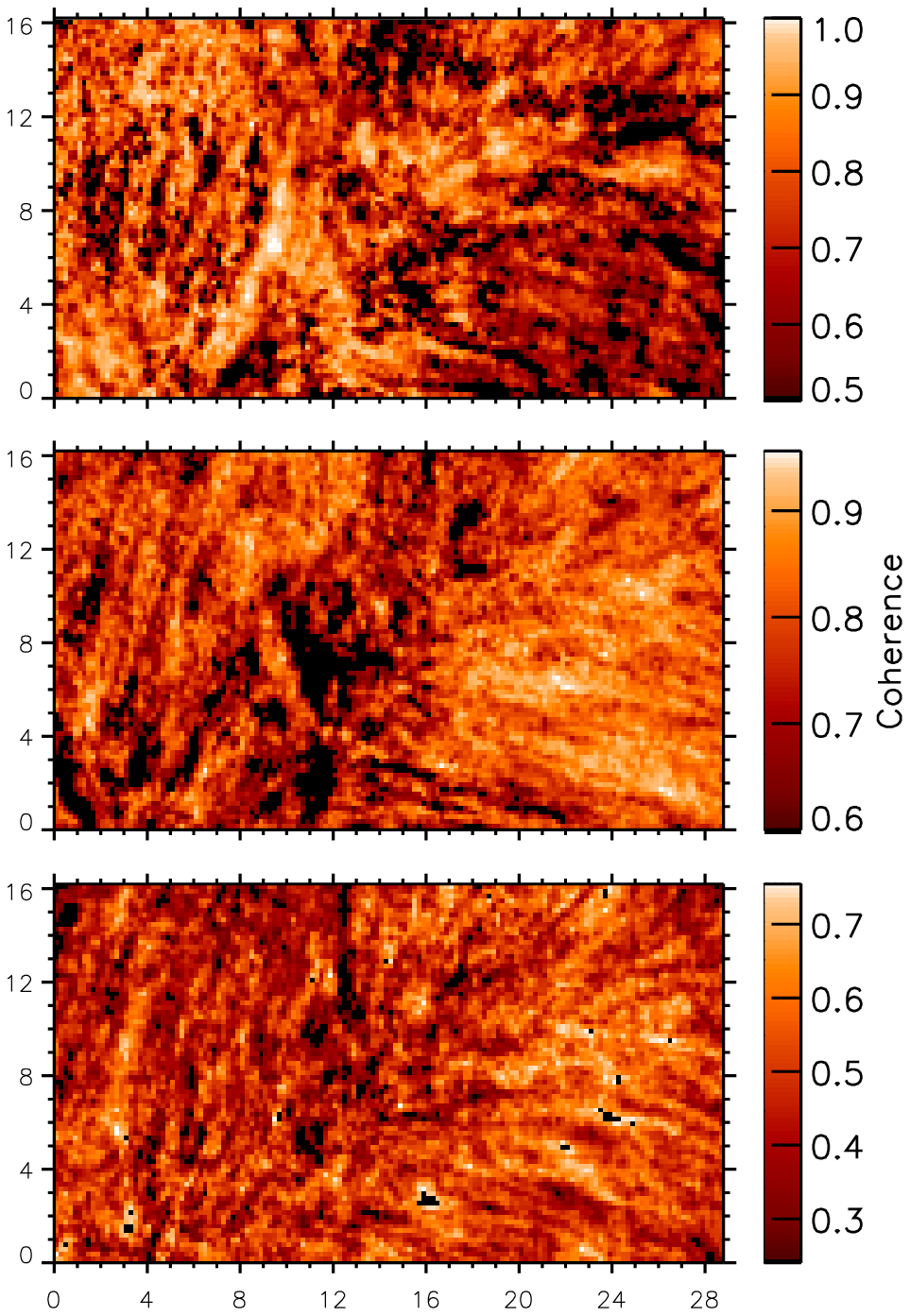}
\includegraphics[scale=0.71]{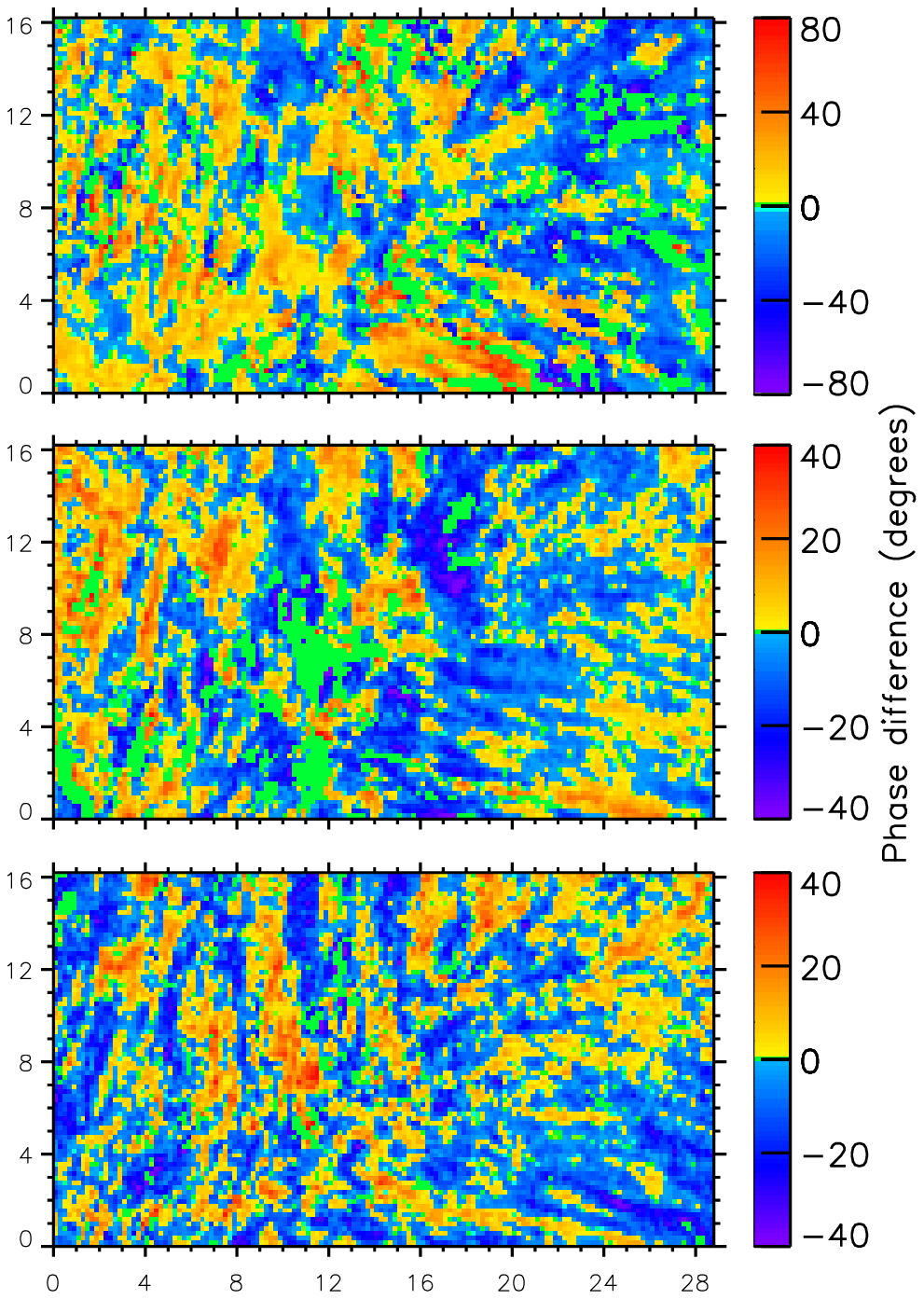}
\end{center}
\caption{Coherence ({\it left}) and phase difference maps ({\it right})
  between Doppler fluctuations at 0.3 and 0.7\,\AA\ from the
  H$\alpha$ center for the network region. Rows from up to down are
  the maps averaged over the selected frequency bands: 0.73\,--\,1.75,
  2.46\,--\,4.16 and 4.93\,--\,8.31\,mHz, respectively. Positive ({\it red})
  and negative ({\it blue}) phase differences correspond to upward and
  downward propagation, respectively. The darkest areas in coherence
  maps correspond to pixels, where coherence does not lie within
  2$\sigma$. The same areas are marked in a
  green color in phase maps, and have been ignored for further
  analysis.}
\label{fig:cohpha_network}
\end{figure*}

\begin{figure*}
\begin{center}
\includegraphics[scale=0.7]{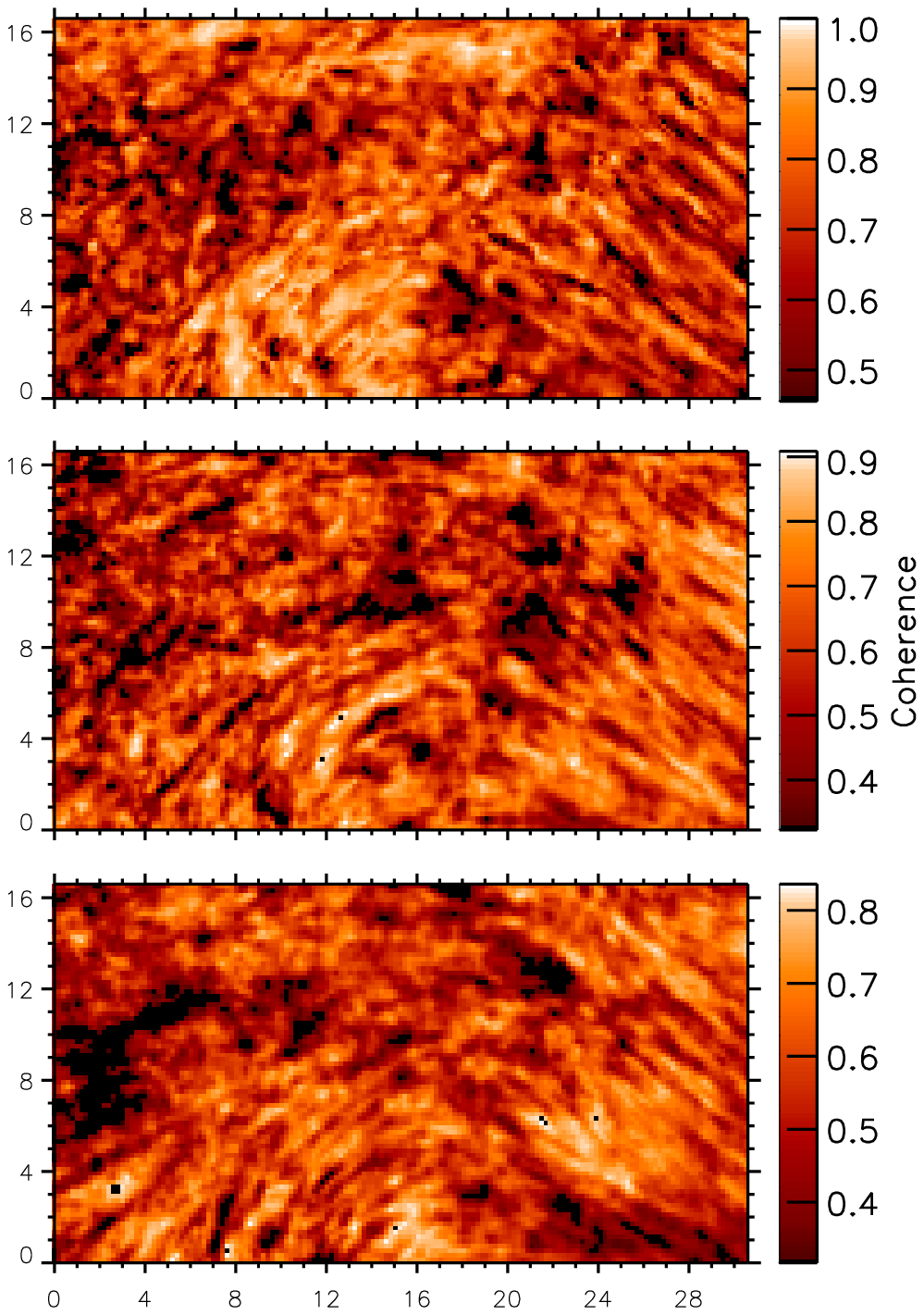}
\includegraphics[scale=0.7]{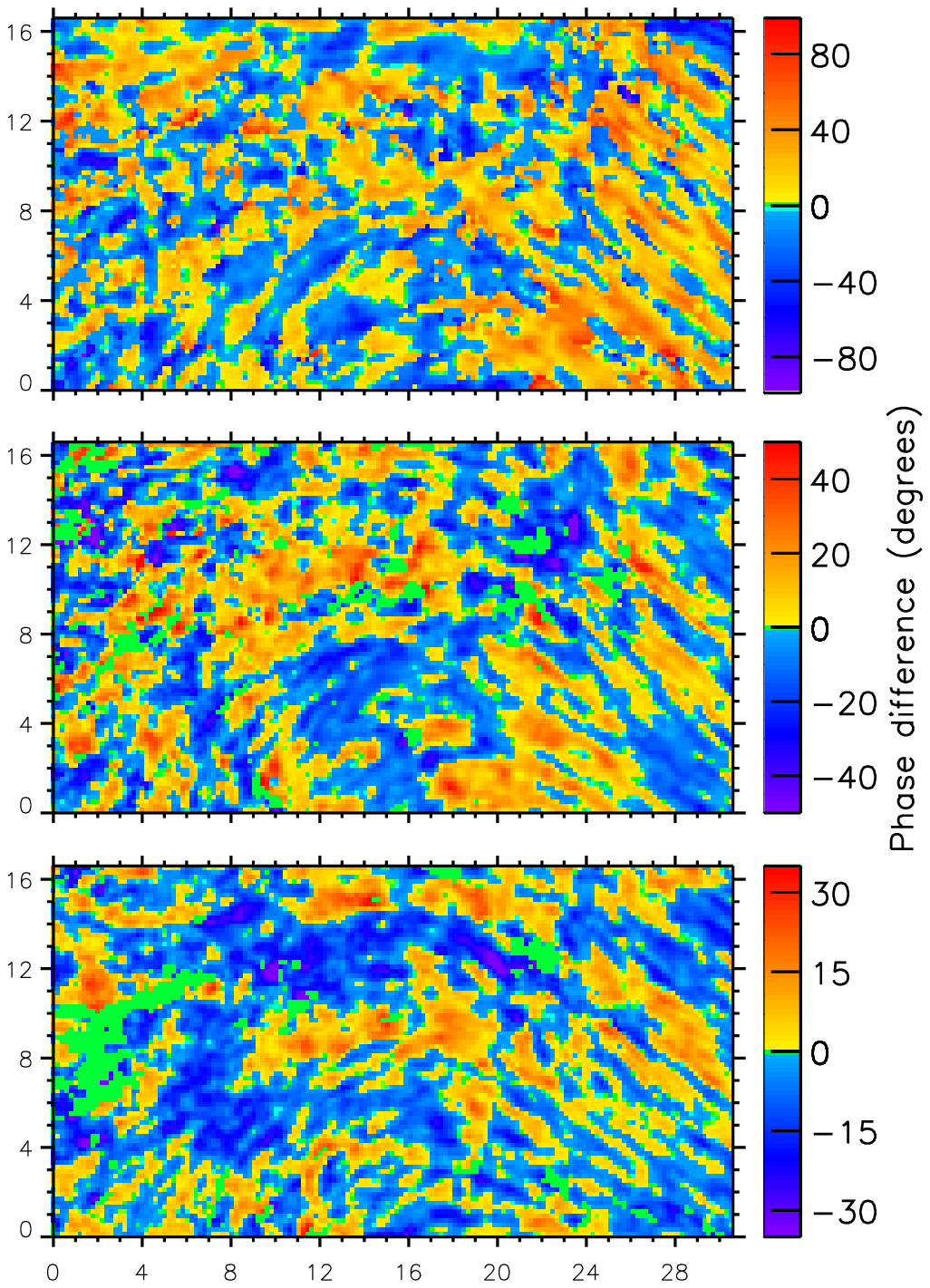}
\end{center}
\caption{The same as Fig.~\ref{fig:cohpha_network}, but for the
  semi-active region.}
\label{fig:cohpha_active}
\end{figure*}

We present the averaged Doppler power, coherence and phase difference
maps over the three frequency bands for the mottles and fibrils
regions in Figures~\ref{fig:2D_fourier},~\ref{fig:cohpha_network}
and~\ref{fig:cohpha_active}. Brighter patches in the power maps
corresponds to strong power. Positive (red) and negative (blue) values
in phase difference maps indicate upward and downward propagation,
respectively.

%--------------------------------------------
%--------------------------------------------

%--------------------------------------------
%------------------------------------------------

%------------------------------------------------
%------------------------------------------------

\vspace{6pt} \noindent {\bf Low frequency range, 0.73\,--\,1.75\,mHz:}
In the mottles region the power spreads from the network to the
structures at the highest atmospheric layer while it is more
concentrated on the network at lower layers. In the fibrils region the
power seems to cover the whole FOV at the highest layer, but the power
occurs primarily on fibrils, especially on the fibril group located
around the lower center part of the FOV. The power also shows granular
distribution in the absence of mottles or fibrils at the lower layers.

The network region, outlined with a large number of bright
  points, is positioned diagonally in the FOV (see the mean intensity
  image for H$\alpha$\,$\pm$\,0.7\,\AA\, in
  Figure~\ref{fig:meanimage}). Here, the phase differences are mostly
  positive (a mean value of about 20$^{\circ}$) with higher
  coherence, showing the existence of upward propagating waves.  In
  the inter-network regions covered by inclined mottles, phase
  differences have both positive and negative values with a mean value
  of about --3$^{\circ}$, indicating the presence of both upward and
  downward propagating waves. Similarly, \cite{gupta13} observed
pronounced upward propagation at magnetic network regions and a
mixture of upward and downward propagation at other locations.
Furthermore, \cite{vecchio07} showed that chromospheric acoustic power
at lower frequencies below the acoustic cut-off, directly propagates
upward from the photosphere in the proximity of the magnetic network
elements.

In the fibrils region, higher coherence is specially seen
in the fibril group located on the lower center part of the FOV, whereas
fibrils located on the right part of the FOV have slightly lower
coherence. In phase difference maps, fibrils on lower center of FOV
have both positive and negative values while on the right of FOV
positive phase differences are more significant in the form of
fibrils. Similarly the contrast in these fibril groups was also
observed on power maps; on fibrils at the lower center part, power is
getting dimmer through the higher atmospheric layers, whereas power is
getting more significant on fibrils at the right part. This increase
in the power can be interpreted as leakage of p-modes along inclined
magnetic fields into the chromosphere \citep{depontieu04, jefferies06,
  vecchio07}. This reveals differences in magnetic topologies of two
fibril groups in the FOV as well.

\vspace{6pt} \noindent {\bf Intermediate frequency range,
  2.46\,--\,4.16\,mHz:} this is centered at 5 minutes. At the highest
atmospheric height, for both observed regions, the power appears to
occur somewhat diffused over the whole FOV with a small tendency to be
less suppressed in areas where the power in low frequency range is
significant. However, at lower atmospheric heights the power
distribution is more pronounced on magnetic fields in the shape of
mottles or fibrils. This is more obvious in the semi-active region.
 
Phase differences are negative with a mean value of 
  about --25$^{\circ}$ in the surrounding network, where the coherence
  has higher value. This area corresponds to the food parts of mottles
  and seems quite dense as a result of overlapping of these parts.
  The negative phase differences here indicate downward reflected
  waves by the inclined magnetic fields. However, in areas far from
  the network, where distances between neighbouring mottles get larger
  (the upper left or the lower right corner parts of FOV), the phase
  differences are mostly positive. This may point to waves which
propagate upward and are not impeded by the inclined magnetic fields
\citep{kontogiannis10}. The downward propagation shows itself in power
maps as well. The power in areas around the magnetic network is
suppressed at chromospheric heights, but it is significant at
photospheric heights. \cite{lawrence10} found power suppression in the
presence of magnetic fields for both G-band and Ca II H-line
oscillations in the frequency range
5.5\,$<$\,f\,$<$\,8.0\,mHz. \cite{kontogiannis10} observed similar
power suppression in 3 and 5 minute ranges at chromospheric heights
but they found it more significant in the 3 minute range.  
  \cite{morton14} reported that the velocity power appears to decrease
  significantly from the chromosphere to the corona, with the power of
  the high-frequency waves decreasing to a much greater degree.  In
our observation, we found that the power suppression exists in mottles
in the 3 and 5 minute ranges although more noticeable at 3
minutes. This phenomena is known as 'magnetic shadow' described
  in the introduction. Using 2D simulations \cite{rosenthal02} showed
  that in regions called magnetic canopy, where the magnetic field is
  significantly inclined to the vertical, waves could be reflected at
  a surface whose altitude is highly variable having consequences on
  the oscillatory processes.

In the 5 minute band the phase differences have mostly negative values
in fibrils on lower center part of the FOV, whereas fibrils on the
right part of the FOV show both positive and negative phase
differences. In the corresponding power maps, while the power
concentrates in fibrils at photospheric heights, it gets weaker and
shows almost uniform distribution with a tendency to be some
significant on areas where a large number of magnetic features are
closely packed at chromospheric heights.
  
\vspace{6pt} \noindent {\bf High frequency range, 4.93\,--\,8.31\,mHz:} 
this is the 3 minute range. It can be noticed that both observed 
regions show similar behavior. The power distribution shows 
suppression in locations of mottles or fibrils at the highest 
atmospheric layer. However, in contrast with this view, the power 
becomes significant on magnetic fields (i.e. in mottles or fibrils) at 
the lower atmospheric heights. 
 
Coherence maps reveal fibrillar structures clearly in the 3
minute range as well as in the 5 minute range. There seems to be
an increase of the coherence in the lower-right area of the mottles
FOV. This area corresponds to the region of less dense
structures. Both positive and negative phase values are seen over the
FOV. Considering the power maps, one can conclude that the wave
reflection for this frequency range is stronger than that for the 5
minute range \citep{vecchio07}.  As a result, the power suppression in
the 3 minute range seems to be stronger than the suppression in the 5
minute range. However, the phase values do not reflect this
situation. This finding might indicate that waves of downward
propagation are overwhelmed by waves of upward
propagation. \cite{dewijn09} reported a similar behaviour for the 5
minute range.
 
In the phase difference maps of the fibrils region, both positive
and negative values are seen in two fibril groups in the FOV, while
negative phase values are noticeable in the fibril-free
region. However, in the power maps from photosphere to chromosphere,
fibrils shows a power suppression while the fibril-free region
shows a power increase. It can be said that positive phase difference
may indicate waves which are not obstructed by the inclined magnetic
fields and travel upward, while downward directed waves may be the
result of the reflection of acoustic oscillations at the inclined
magnetic fields \citep{kontogiannis10}. On the other hand, this view
does not explain the negative phase values, which occurs in the
fibril-free region.  On the contrary, there is also a power increase
in this region upward to the chromosphere. This could indicate
downward directed and damped oscillations. The same area shows
positive phase values and a diffuse power increase in the
chromospheric heights in the 5 minute band. It is questionable if this
is a result of wave refraction.

\begin{figure*} 
\begin{center} 
\includegraphics[scale=0.8]{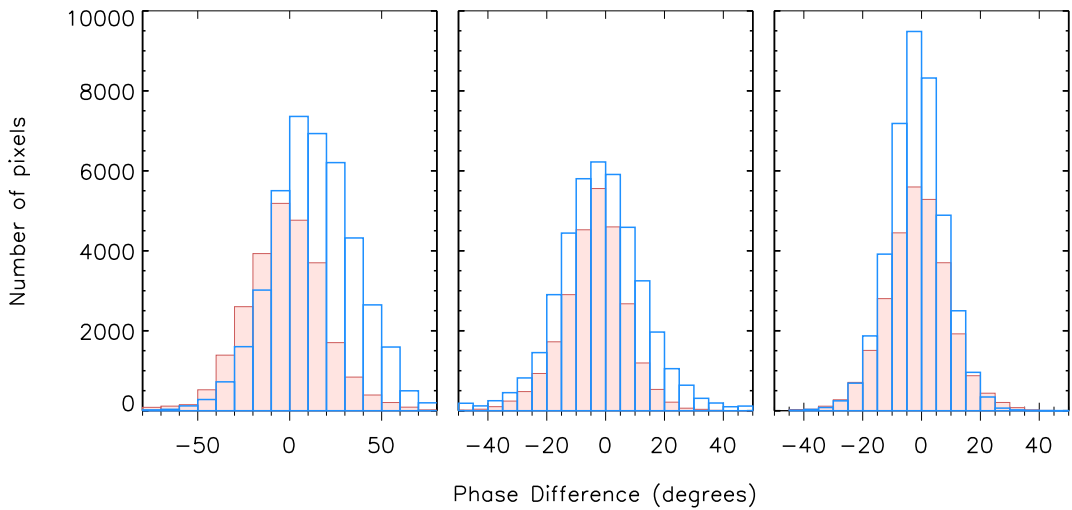} 
\end{center} 
\caption{Distributions of phase differences for mottle regions ({\it red 
    filled bars}) and fibril regions ({\it blue unfilled bars}). Panels 
  from left to right show low, intermediate and high frequency 
  ranges, respectively.} 
\label{fig:histograms} 
\end{figure*}

Following dark elongated fine structures such as mottles and fibrils 
during their life time in H$\alpha$ is a complex task since their 
shape and length change continuously and their contrast is not 
sufficient to outline these structures. For the definition of their 
areas in the FOV we used the time averaged intensity image at 0.3~\AA\ 
from the line center and masked the pixels, where values lie under the 
mean value of the image.  We show in Figure~\ref{fig:histograms} the 
distributions of phase differences in the three frequency bands for 
both structures (mottles and fibrils) selected by masks. In the low 
frequency range, while the phase values for the mottles exhibit a 
Gaussian distribution, the phases for the fibrils lie predominantly in 
the positive region. In the intermediate frequency range, the phase 
values for the mottles lie mainly in the negative region, while in the 
high frequency range they show a relatively symmetrically distribution 
around zero confirming the results derived from the phase maps. The 
phase values for the fibrils in the 5 and 3 minute ranges are also 
symmetrically distributed around zero, showing a slightly higher 
number of negative values, but there is no obvious dominance. 

%-------------------------------------------- 
%--------------------------------------------  
 
\section{Conclusion} 
\cite{dewijn09} investigated the oscillations in a plage and expressed
that the 5 minute oscillations can only propagate along inclined
magnetic field lines and not in the central region of the plage, where
the magnetic field is expected to be vertical. However, in the present
study, a power increase is seen in the chromospheric heights on
network bright points in the 5 minute band. \cite{roberts83}
  proposed that radiative losses in thin flux tubes can lead to a
  significant reduction of the cutoff frequency. This mechanism
  enables propagation of the 5-minute oscillations to the chromosphere
  in the vertical magnetic structures. Afterwards, \cite{centeno06,
    centeno09} and \cite{khomenko08} stated that if the radiative
relaxation time is short enough, then the acoustic cut-off frequency
would decrease to lower values and this would allow the vertical
propagation of 5 minute waves to the chromosphere. \cite{heggland11}
found in their simulations 5 minute long-period oscillations to be
dominant in strong and inclined magnetic field, while the 3 minute
oscillations to be dominant in areas with weak or vertical magnetic
fields. \cite{nutto12} connected the slow acoustic mode power observed
on magnetic network elements to mode conversion. In the present study,
the power of the 3 minute oscillations gets stronger with increasing
atmospheric heights in regions where the magnetic field is weak. But
it is difficult to say that the 5 minute waves propagate along
inclined magnetic fields. On the contrary to this, we found for both
regions in areas of inclined magnetic fields a power suppression,
which is stronger in the 3 minute band, confirming the results of
\cite{kontogiannis10}. A power increase in inclined magnetic fields
through higher atmospheric heights is seen for the low frequency
range, which may be a signature of leakage of p-modes 
  \citep{depontieu04, jefferies06, vecchio07}. In the same frequency
range, our investigations revealed the presence of propagating waves
in the magnetic network, which was also reported by \cite{gupta13}. In
relation with the intermediate frequency range, a weak power
suppression was found in regions of inclined magnetic fields for both
FOVs showing negative phase differences and indicating downward
reflected waves. A stronger suppression was observed for the high
frequency range, while the phase values did not reflect this
case. This may be due to a overwhelming of downward propagating
signals by upward propagating signals. One of the interesting results
of this study is the negative phase values in the 3 minute band and
the positive phase values in the 5 minute band, which occurs for the
same area in a fibril-free region. A possible explanation for the
negative phase values could be downward directed and damped
oscillations since the power decreases towards lower heights.

\section*{Acknowledgments} 
We would like to thank the referee for valuable comments and
suggestions. We are also indebted to Franz Kneer for helpful comments
and suggestions.  A. G\"ultekin recognizes a support from Scientific
Research Projects Coordination Unit of Istanbul University. Project
number UDP-24161.  This work was partially supported by Deut\-sche
For\-schungs\-ge\-mein\-schaft through grant Kn 152/26-1. The Vacuum
Tower Telescope is operated by the Kiepenheuer-Institut f\"ur
Sonnenphysik, Freiburg, at the Spanish Observatorio del Teide of the
Instituto de Astrof$\acute{\i}$sica de Canarias.

\end{document}